\documentstyle[aps,twocolumn]{revtex}
\draft
\def\ba{\begin{eqnarray}}
\def\ea{\end{eqnarray}}

\begin{document}
\preprint{HD-TVP-94-12}

\title{Random matrix theory for CPA: Generalization
of Wegner's $n$--orbital model}
\author{Peter Neu$^1$ and Roland Speicher$^2\dagger$}
\address{$^1$ Institut f\"ur Theoretische Physik, Universit\"at Heidelberg,
Philosophenweg 19, 69120 Heidelberg, Germany}
\address{$^2$ Institut f\"ur Angewandte Mathematik, Universit\"at Heidelberg,
Im Neuenheimer Feld 294, 69120 Heidelberg, Germany}
\date{August 26, 1994}
\maketitle
\begin{abstract}
 We introduce a generalization of Wegner's $n$-orbital model for
the description of randomly disordered systems by replacing his
ensemble of Gaussian random matrices by an ensemble
of randomly rotated matrices.
We
calculate the one- and two-particle Green's functions and
the conductivity exactly in the limit $n\to\infty$.
Our solution solves the CPA-equation of the $(n=1)$-Anderson
model for arbitrarily distributed  disorder.
We show how the Lloyd model is included in our model.
\end{abstract}

\pacs{PACS numbers: 71.10.+x, 05.60.+w, 71.55.-i}

\narrowtext

 The treatment of physical systems with disorder presents one of the
great challenges in statistical physics. Realistic multi-site
models, like the Anderson model \cite{And},
are in general unsolvable. Exact
calculations are only possible in one pathological special case -
namely for the Lloyd model \cite{Llo}
with Cauchy-distributed disorder. Otherwise
one has to use approximation techniques which reduce the multi-site to
single-site models and treat them self-consistently. The most prominent
of these methods is the so-called coherent potential approximation (CPA),
see, e.g., \cite{LGP}.

Wegner \cite{Weg} introduced a generalization of the Anderson model
by putting $n$ electronic states at each site and describing the
disorder by Gaussian random matrices in these electronic states.
Whereas for $n=1$ this reduces to the usual (unsolvable) Anderson
model, the opposite limit $n\to\infty$ becomes exactly solvable.
Interestingly, the solution of this multi-site model coincides with a
special CPA-solution.

In this letter we introduce a generalization of Wegner's model by
replacing his ensemble of Gaussian random matrices by a more general
ensemble of random matrices - and thus allowing arbitrarily
distributed disorder. Nevertheless - by using recent results of
Voiculescu \cite{Void,VDN} and Speicher \cite{Spez,NSp}
on the mathematical concept
of ``freeness" - we are still able to calculate the Green's functions of
this model in the limit $n\to\infty$ exactly. Again our solution
solves the CPA-equation of the $(n=1)$-Anderson model. Hence our
model can be regarded as a rigorous mean-field model for CPA for
arbitrarily distributed disorder.

Let us first recall Wegner's model. He considers a $d$-dimensional
lattice where at each site $r$ there are $n$ electronic levels
$|r\alpha\rangle$ numbered by $\alpha=1,\dots,n$. The interaction is
governed by a Hamiltonian of the form
\ba  \label{einsa}
 H=H_0+H_1,
\ea
where $H_0$ is deterministic, translational-invariant, and
diagonal in the electronic levels,
\ba  \label{einsb}
 H_0=\sum_{r,r',\alpha} v_{r-r'}
|r\alpha\rangle\langle r'\alpha|,
\ea
and $H_1$ describes the site-diagonal disorder
\ba  \label{einsc}
H_1=\sum_{r,\alpha,\beta} \frac 1{\sqrt n} f_r^{\alpha\beta}
|r\alpha\rangle\langle r\beta|.
\ea
There, $f_r=\big((1/\sqrt n)f_r^{\alpha\beta}\big)_{\alpha,\beta=1}^n$
are Gaussian random matrices and the entries of $f_r$ and of $f_{r'}$ are
independent for $r\not= r'$. (This is the site-diagonal model of Wegner,
we will not treat his  ``local gauge invariant" model here.)
By using techniques for calculating moments of Gaussian random matrices
\cite{Wig,Arn}, Wegner was able to calculate the one- and two-particle
Green's function in the limit $n\to\infty$.

To explain our generalization let us diagonalize the random matrices
$f_r$ in the form
 $f_r=u_r\hat f_r u_r^*$.
Thus the ensemble of the $f_r$ is determined by an ensemble of diagonal
matrices $\hat f_r$ and an ensemble of unitary matrices $u_r$.
In the limit $n\to\infty$, Wegner's original formulation is recovered
if the diagonal $\hat f_r$ are taken as deterministic matrices having
Wigner's semi-circle \cite{Wig} as eigenvalue distribution and
if the $u_r$ are random
unitary matrices, given by
the canonical invariant or Haar measure on $U(n)$, such that $u_r$
and $u_{r'}$ are chosen independently from $U(n)$ for $r\neq r'$.
Since this ensemble is invariant
under independent rotations at different sites
we may also replace the different $\hat f_r$ by one single (not
necessarily diagonal) $f$, i.e. we have $f_r=u_rfu_r^*$ with the
ensemble of $u_r$ as stated above.

The advantage of this reformulation of Wegner's model is that now
a generalization is obvious: We are a priori totally free in the
choice for $f$.
Thus our model is given by the following Hamiltonian
\ba   \label{zwei}
H=\sum_{r,r',\alpha}v_{r-r'} |r\alpha\rangle\langle r'\alpha | +
\sum_{r,\alpha,\beta}(u_rfu_r^*)_{
\alpha,\beta}  |r\alpha\rangle\langle r\beta |,
\ea
where $f$ is a deterministic hermitian $n\times n$-matrix and $u_r\in U(n)$
are random unitary matrices, chosen independently for different sites.
This means that we act at each site $r$ with a
copy $f_r:=u_rfu_r^*$ of the given
operator $f$, but that the basis for $f_r$ and the basis for $f_{r'}$
are rotated randomly against each other for all pairs of different
sites $r\neq r'$.

The possibility for treating this model in the limit $n\to\infty$ arises
from the important observation of Voiculescu \cite{Void} (see also
\cite{Spee}) that there is a connection with his concept of free random
variables \cite{Voie,VDN,Sped}:
Denote by $\langle\dots\rangle_{\rm ens}$ the average over our ensemble of
random unitary matrices and let
\ba
 \langle\dots\rangle := \left\langle \frac {1}{n}
\sum_{\alpha=1}^n\langle\alpha |\dots |\alpha\rangle\right\rangle_{\rm ens}
\ea
count the averaged eigenvalue distribution of our $n\times n$-matrices.
Then we have for all $m\in $ {\bf N}  and for all polynomials $p_1,\dots,p_m$
with $\langle p_i(f)\rangle = 0$ ($i=1,\dots,m$) in the limit $n\to\infty$ that
also
\ba   \label{drei}
 \langle p_1(f_{r(1)})p_2(f_{r(2)})\dots p_m(f_{r(m)})\rangle = 0
\ea
for all sequences of indices $r(1),\dots,r(m)$, where all consecutive
indices are different, i.e. where $r(i)\neq r(i+1)$ for all $i=1,\dots,m-1$.
An example of (\ref{drei}) is
\ba   \label{viera}
 \langle p_1(f_1)p_2(f_2)p_3(f_1)p_4(f_2)\rangle = 0,
\ea
whereas for sequences with coinciding neighbouring indices one gets
non-vanishing results like
\ba   \label{vierb}
 \langle p_1(f_1)p_2(f_2)p_3(f_2)p_4(f_1)\rangle  &=& \\
\langle p_1(f_1)p_4(f_1)\rangle \langle p_2(f_2)p_3(f_2)\rangle
 &=& \langle p_1(f)p_4(f)\rangle \langle p_2(f)p_3(f)\rangle. \nonumber
\ea
The property (\ref{drei}) - in the mathematical literature recently
introduced by Voiculescu \cite{Voie,VDN}
under the name of  ``freeness"  - allows,
as in Wegner's case, to calculate
all mixed moments of the matrices $f_r$ (like, e.g., in (\ref{vierb})) and
thus to derive exact expressions for the Green's functions of
our model in the limit $n\to\infty$. Whereas usually one gets
an infinite hierarchy of equations for averaged quantities, it is
exactly the property of freeness (\ref{drei}) which closes our equations.
For an effective handling of the calculations one needs the
 $R$-transform machinery of Voiculescu \cite{Voiz,VDN} and the concept
of non-crossing cumulants of Speicher \cite{Spez,NSp}. The concrete
calculations will be published elsewhere, here we only want to give
the results.

The most important quantity is (the diagonal part of) the
one-particle Green's function (1PG)
\ba  \label{funf}
 G(z) := \left\langle\frac 1n\sum_{\alpha=1}^n\left\langle
r\alpha\left|\frac1{z-H}\right|
r\alpha\right\rangle\right\rangle_{\rm ens}
\ea
whose spectral function yields the density of states.
Note that, due to the translation-invariance of our Hamiltonian, $G(z)$ is
independent of $r$ and hence the local and the global
density of states coincide.
If we denote similarily by
$G_0$ and $G_1$ the 1PG of the deterministic and the disorder part
of $H$, respectively, i.e.
\ba  \label{sechsa}
 G_0(z)&=&\left\langle r\left|\frac {1}{z-H_0}\right|r\right\rangle \\
G_1(z)&=&\left\langle
\frac {1}{n}\sum_{\alpha=1}^n\left\langle \alpha\left|\frac{1}{z-f_r}\right|
\alpha\right\rangle\right\rangle_{\rm ens}\nonumber\\
&=&\frac {1}{n}\sum_{\alpha=1}^n\left\langle \alpha\left|\frac{1}{z-f}\right|
\alpha\right\rangle,
\ea
then the crucial property (\ref{drei}) allows us to derive the following
self-consistent equation for $G(z)$:
\ba  \label{siebena}
 G(z) = G_0\Big( z-R_1\big(G(z)\big)\Big) ,
\ea
where $R_1(z)$ is a kind of self-consistent self-energy for
$H_1$, namely it is defined by
\ba  \label{siebenb}
 G_1(z) = \frac {1}{z-R_1\big( G_1(z)\big)}
\ea
with $R_1(0)=0$.
In the special case of a semi-circle eigenvalue
distribution for $f$, $G_1$ is
given by
\ba   \label{acht}
 G_1(z) = \frac{z-\sqrt{z^2-4\sigma^2}}{2\sigma^2}=\frac {1}{z-\sigma^2G_1(z)}
\ea
showing that $R_1(z)=\sigma^2z$. In this case our formula (\ref{siebena})
reduces to Wegner's result \cite{Weg,KP}
\ba   \label{neun}
 G(z) = G_0\Big( z-\sigma^2G(z)\Big).
\ea
Given $G(z)$ and $R_1(z)$, one can derive linear self-consistent
equations for the one-particle and two-particle Green's functions.
Again, the freeness property (\ref{drei}) is the essential ingredient
for closing the equations.
For the Fourier-transform of the one-particle Green's function one
obtains
\ba  \label{zehna}
 \tilde G(q;z) = \frac {1}{z-v(q)-R_1\big( G(z)\big)}
 \ea
with $v_r = \int_q v(q) e^{iqr}$ where
$ \int_q = \frac{{\cal V}}{(2\pi)^d}\int_{1BZ} d^dq$,
 ${\cal V}$ being the volume of
the first Brillouin zone ($1BZ$).
For $R_1(z)=\sigma^2z$ this reduces again to Wegner's result
\cite{Weg,KP}.

The two-particle Green's function determines the conductivity as
a velocity-velocity correlation function. Although in general the
two-particle Green's function contains an additional term,
one obtains by the symmetry argument
$v_{-r}=v_r$ that
only the product of the one-particle Green's functions
contribute. Thus
one finds for zero temperature and in the dc-limit $\omega
\to 0$
\ba   \label{elf}
\sigma(\omega,0) = \frac {2\pi e^2 B}{{\cal V}}\, \varrho^2(R_1(G);E_F)
\ea
where
\ba
 B :=
\int_q\frac{|\nabla v(q)|^2}
{| v(q)-\zeta(E_F+\omega+i0^+)|^2
| v(q)-\zeta(E_F+i0^+)|^2}  \nonumber
\ea
with $\zeta(z)=z-R_1(G(z))$ and the spectral function
\ba
 \varrho(R_1(G);E) \,=\, -\frac {1}{\pi}\, \mbox{Im}\,
R_1\big(G(z=E+i0^+)\big).
\ea
For $R_1(z)=\sigma^2z$ this reduces to Wegner's result in a form given
by Khorunzhy and Pastur \cite{KP}.

The generality of our solution may be seen from the fact that it
includes also the Lloyd model as a special case. Namely,
choose a Cauchy-distribution with parameter $\gamma$
as eigenvalue distribution for $f$. Then
\ba  \label{zwolf}
 G_1(z) = \frac {1}{\pi}\int_{-\infty}^{+\infty}\frac{\gamma}{\gamma^2+t^2}
\frac {1}{z-t}dt=\frac {1}{z-i\gamma},
\ea
hence $R_1(z)=i\gamma$.
This shows that the Cauchy-distribution behaves in all relevant aspects
like an imaginary $\delta$-distribution. For $\delta$-distributions,
however, there is no difference between our model and the original
Anderson Hamiltonian
(see also \cite{Maa,BV}). Since the Anderson model with Cauchy-distributed
disorder is nothing else but the Lloyd model, the latter is included in
our investigations.
In particular, we recover
from (\ref{siebena}) the 1PG of the Lloyd model \cite{Llo}
\ba  \label{dreizehn}
 G(z) = G_0\big( z-i\gamma\big).
\ea

One surprising feature of the Wegner model is that its solution
coincides with a special CPA-solution. This generalizes also to
our model.
In general, the CPA-solution
for the Anderson model with
single-site random variable $X$ is given by the two equations
\ba  \label{vierzehna}
 G(z) = G_0\big( z-\Sigma(z)\big)
 \ea
and
\ba  \label{vierzehnb}
\left\langle
\frac{X-\Sigma(z)}{1-(X-\Sigma(z))G(z)}\right\rangle = 0.
\ea
The first of these equations coincides with our solution (\ref{siebena})
if we identify
$\Sigma(z)=R_1(G(z))$. By using the equivalent
form of (\ref{siebenb}), namely
\ba  \label{funfzehn}
 G_1\big(R_1(z)+z^{-1}\big) = z,
\ea
it can be checked
that our solution fulfills also (\ref{vierzehnb}) if we choose as distribution
for $X$ the eigenvalue distribution of $f$.
Note that we specify our
model rigorously in the
beginning and that we are able to calculate all quantities without
any further approximation. Thus our multi-site model is a
rigorous mean-field model
for the usual single-site CPA. One should also note
that our previous remarks about the Lloyd model can now be taken as an
explanation for the well-known fact that CPA is exact for the Lloyd model.

As an instructive example of our formalism let us consider the
one-dimensional lattice with nearest-neighbour interaction
and binary site-diagonal disorder for the special case
$v=\sigma=1$, where $2v$ is
the half-band-width and $\sigma^2$ is the variance of the disorder.
Then we have $G_0(z)=1/\sqrt{z^2-1}$ and $R_1(z)=(\sqrt{1+z^2}-1)/
2z$, which yields as a solution of $(\ref{siebena})$
\ba
 G(z) = \frac{8}{6\sqrt{z^2-2}+2z}.
\ea
This provides
\ba
 \varrho(R_1(G);E) = \frac {1}{4\pi} \sqrt{2-E^2}\ \Theta(\sqrt 2 -E),
\ea
i.e. together with (\ref{elf}) a finite conductivity everywhere inside
the band $[-\sqrt 2, \sqrt 2]$.

Since we can prescribe in our model an arbitrary distribution for $f$
- or for $X$ in the CPA-formulation -, it is clear that our formalism
is capable of covering a lot of quite different examples, like, e.g.,
a superposition of binary noise or, more general, the recently investigated
$q$-noise
\cite{NSpz}. Furthermore, our description using the theory of free
random variables and the notion of non-crossing cumulants
allows a straightforward generalization to the case of dynamical
disorder and thus promises to give a
rigorous model for dynamical CPA. These
subjects will be
pursued further in forthcoming investigations.

We acknowledge helpful  discussions with  Franz Wegner and Petr Chvosta.
This work was supported by the Deutsche Forschungsgemeinschaft (R.S.).


\begin{references}

\bibitem[\dagger] {i} Supported by a  fellowship from the DFG


\bibitem  {And}
 P.W. Anderson, Phys. Rev. {\bf 109}, 1492 (1958)

\bibitem {Llo}
P. Lloyd, J. Phys. {\bf C} {\bf 2}, 1717 (1969)

\bibitem {LGP} I.M. Lifshitz, S.A. Gredeskul, and
L.A. Pastur, {\it Introduction to the theory of disordered systems}
(Wiley, New York, 1988)

\bibitem {Weg} F. Wegner, Phys. Rev. {\bf B} {\bf 19}, 783 (1979)

\bibitem {Void}  D. Voiculescu, Invent. math. {\bf 104}, 201 (1991)

\bibitem  {VDN} D. Voiculescu, K. Dykema, and A. Nica,
{\it Free Random Variables} (AMS 1992)

\bibitem  {Spez} R. Speicher, Math. Ann. {\bf 298}, 611 (1994)

\bibitem  {NSp} P. Neu and R. Speicher, Z. Phys. {\bf B} {\bf 92}, 399 (1993)

\bibitem  {Wig}  E.P. Wigner, Ann. Math. {\bf 62}, 548 (1955);
{\bf 67}, 325 (1958)

\bibitem  {Arn} L. Arnold, J. Math. Anal. Appl. {\bf 20}, 262 (1967)

\bibitem  {Spee}  R. Speicher, RIMS {\bf 29}, 731 (1993)

\bibitem  {Sped} R. Speicher, Prob. Th. Rel. Fields {\bf 84},
141 (1990)

\bibitem  {Voie}  D. Voiculescu,
Lecture Notes in Mathematics {\bf 1132}, 556 (1985)

\bibitem  {Voiz}  D. Voiculescu, J. Funct. Anal. {\bf 66}, 323 (1986)

\bibitem {KP} A.M. Khorunzhy and L.A. Pastur, Commun.
Math. Phys. {\bf 153}, 605 (1993)

\bibitem  {Maa}  H. Maassen, J. Funct. Anal. {\bf 106}, 409 (1992)

\bibitem  {BV}   H. Bercovici and D. Voiculescu, Indiana U.
Math. J. {\bf 42}, 733 (1993)

\bibitem {NSpz} P. Neu and R. Speicher, Z. Phys. {\bf B} {\bf 95}, 101 (1994)


\end{references}
\end{document}